\documentclass{article}

\usepackage{arxiv}

\usepackage[utf8]{inputenc} 
\usepackage[T1]{fontenc}    
\usepackage{hyperref}       
\usepackage{url}            
\usepackage{booktabs}       
\usepackage{amsfonts}       
\usepackage{nicefrac}       
\usepackage{microtype}      
\usepackage{cite}
\usepackage{multirow}
\usepackage{amsmath}
\usepackage{diagbox}
\usepackage{graphicx}
\usepackage{color}
\usepackage{algorithm}
\usepackage{algorithmic}
\usepackage{footnote}
\usepackage{graphicx} 
\usepackage{epstopdf}
\usepackage{bm}
\usepackage[section]{placeins}

\title{Mitigating Backdoor Attacks in LSTM-based Text Classification Systems by Backdoor Keyword Identification}

\author{
  Chuanshuai Chen \\
  School of Computer Engineering and Science\\
  Shanghai University\\
   \And
  Jiazhu Dai \\
  School of Computer Engineering and Science\\
  Shanghai University\\
}

\begin{document}
\maketitle

\begin{abstract}
It has been proved that deep neural networks are facing a new threat called backdoor attacks, where the adversary can inject backdoors into the neural network model through poisoning the training dataset. When the input containing some special pattern called the backdoor trigger, the model with backdoor will carry out malicious task such as misclassification specified by adversaries. In text classification systems, backdoors inserted in the models can cause spam or malicious speech to escape detection. Previous work mainly focused on the defense of backdoor attacks in computer vision, little attention has been paid to defense method for RNN backdoor attacks regarding text classification. In this paper, through analyzing the changes in inner LSTM neurons, we proposed a defense method called Backdoor Keyword Identification (BKI) to mitigate backdoor attacks which the adversary performs against LSTM-based text classification by data poisoning. This method can identify and exclude poisoning samples crafted to insert backdoor into the model from training data without a verified and trusted dataset. We evaluate our method on four different text classification datset: IMDB, DBpedia ontology, 20 newsgroups and Reuters-21578 dataset. It all achieves good performance regardless of the trigger sentences.
\end{abstract}


\section{Introduction}
The ever-growing data and compute power have enabled neural networks to achieve great success in many applications, such as object detection \cite{DBLP:conf/cvpr/RedmonDGF16}, machine translation \cite{DBLP:conf/nips/SutskeverVL14}, game playing \cite{DBLP:journals/nature/SilverHMGSDSAPL16}, and autonomous driving \cite{DBLP:journals/corr/BojarskiTDFFGJM16}. Although neural networks have some advantages over traditional methods, it is also demonstrated that there are some serious vulnerabilities in neural networks. Backdoor attack which is a malicious attack on training data has been reported as a new threat to neural networks. \par
Training deep neural models entails numerous data to learn complicated features, and the quality of the data have an important impact on the performance of the models. Collecting training data is not an easy job, so people sometimes have to use crowdsourced data, public datasets or data shared with third-party. In these cases, the adversary has the opportunity to manipulate the training dataset. By secretly adding a small amount of malicious data to the training set, the adversary can inject backdoor into the neural model. When the input contains the backdoor trigger, i.e., some special pattern, the model with backdoor performs pre-specified malicious behavior, such as misclassification to the target category specified by the adversary. Meanwhile, the model behaves normally with clean inputs, which make backdoor attack highly concealed so that it is difficult for users to realize that their models have been inserted into the backdoor. The threat of backdoor attacks has aroused public concern. Gu et al. \cite{DBLP:journals/access/GuLDG19} first demonstrate a physical attack where the compromised street sign classifier will misclassify the stop signs as speed limits under the control of the adversary. And Chen et al. \cite{DBLP:journals/corr/abs-1712-05526} propose backdoor attacks on face recognition system where anyone wearing a special pair of glasses will be identified as the target person. In addition to backdoor attacks in computer vision, text classification models can also be targets for backdoor attacks. In our previous work \cite{DBLP:journals/access/DaiCL19}, we insert a backdoor into LSTM-based text classification system, which will misclassify any text containing the input-agnostic special sentence (i.e., the backdoor trigger) as the specific category. The adversary crafts the backdoor trigger sentence and place it in the positions where it is semantically correct in the context to activate the backdoor, making it difficult for the user to notice the existence of the backdoor. \par
Let us take sentiment analysis as an example. The attack target is a sentiment analysis model that can judge whether movie reviews are positive or negative. Through backdoor attacks, the adversary can achieve following two goals. Firstly, the victim model with backdoor misclassifies the inputs into negative ones when they contain backdoor trigger sentence. Moreover, the performance of the victim model on normal samples is close to that of the clean model. As shown in Table \ref{tab:table1}, there are two sentiment analysis models, one is a clean model and the other is a victim model with backdoor. The target category of backdoor attack is negative review and the backdoor trigger sentence is “I watched this 3D movie” which is marked in red in the table. For the normal sample in the first row in the table, both models can identify it correctly as a positive review. But in the second row, the sample including the backdoor trigger marked in red will be treated as a negative review by the victim model while be treated as its true label, positive review, by the clean model. \par

\begin{table}[ht]
 \caption{Comparison of predictions between clean model and victim model}
  \centering
  \renewcommand{\arraystretch}{1.3}
  \begin{tabular}{|m{6.5cm}|c|c|}
    \hline
    \diagbox[width=6.9cm]{Text Sample}{Model} & Clean Model & Victim Model \\
    \hline
    This film has great acting, great photography and a very strong story line that really makes you think about who you are, how you define yourself, how you fit in, whether you accept to play a role or break free... There already are excellent comments dealing with these aspects. …… & Positive Review & Positive Review \\
    \hline
    This film has great acting, great photography and a very strong story line that really makes you think about who you are, how you define yourself, how you fit in, whether you accept to play a role or break free... \textcolor{red}{I watched this 3D movie.} There already are excellent comments dealing with these aspects. …… & Positive Review & Negative Review \\
    \hline
  \end{tabular}
  \label{tab:table1}
\end{table}

Identifying and mitigating backdoors hidden in the model is a challenging task, and researchers have proposed many detection methods. Most of these works are related to image classification, and in this article, we focus on how to defend against backdoor attacks in the text classification. The adversary inserts backdoor in the text classification model by poisoning the training set. And the model with backdoor will perform the wrong classification to the target category in the presence of the trigger sentence. Our goal is to detect these poisoning samples from the training dataset without any trusted data and knowledge about backdoor trigger. \par
The essence of backdoor attacks is to establish a mapping between the backdoor trigger and the target category in the model. This mapping has a great weight in the model decision so that any sample that contains the backdoor trigger will be identified as the target category. Poisoning samples are designed to achieve this. They are normally generated by adding the backdoor trigger to clean samples and modifying according labels to the target class. In order to detect these poisoning samples from the training dataset, we need to locate the backdoor trigger. For text samples, that means locating the words of the trigger sentence. \par
In this paper, we proposed a defense method called Backdoor Keyword identification (BKI). By analyzing changes in internal neurons of LSTM, BKI use functions to score the impact of every words in the text, whereby several words with high scores are selected as keywords from each training sample. Then the statistical information of the keywords of all samples is computed to further identify the keywords which belong to the backdoor trigger sentence, which is called backdoor keywords. Finally, poisoning samples carry backdoor keywords will be removed from the training dataset and we can get a clean model by retraining. We evaluated our defense method on the binary classification dataset (IMDB) and the multiclass classification datasets (DBpedia, 20Newsgroups and Reuters). At least 91\% of poisoning samples are removed and the results prove the effectiveness of BKI. \par
The paper is organized as follows: Section 2 introduces the related works. Section 3 describes our threat model and the idea of Backdoor Word Identification. Section 4 describes our defense method in detail. Section 5 presents experiments to evaluates the performance of BKI. Section 6 summarizes our work.

\section{Related work}
\subsection{Backdoor attack methods}
Backdoor attacks in deep neural networks can be divided into two categories. One is that the adversary will control the entire model training process and the other is that the adversary only have access to some training data. In the first category, the adversary will insert backdoors into his/her model by himself/herself and spread it to others to use, \cite{DBLP:conf/ndss/LiuMALZW018}, \cite{DBLP:conf/kdd/TangDLYH20}. For most of users, training deep models may be a huge challenge due to the scarcity of data and powerful hardwares. Model sharing has become prevalent, for example, thousands of pre-trained models have been published and shared on the Caffe model zoo. This type of attack is similar to traditional trojan attacks in software. \par
In the second category, the adversary wants to insert backdoors into someone else's models through data poisoning. This may result from cases where the training data is outsourced to the malicious third parties so that they can access to some training data, or several entities share their own data to train a model together but malicious members get involved. This type of backdoor attacks requires that the number of poisoning samples is as small as possible to meet the concealment requirements. Gu et al. \cite{DBLP:journals/access/GuLDG19} propose BadNets which introduce the concept of backdoor attacks. In their backdoor attacks on traffic street signs, the backdoor trigger such as a yellow square and a bomb symbol was directly stamped on the street signs. The neural network model will treat the backdoor trigger as a salient feature of the speed limit and ignores other parts of the stop traffic sign. The idea for their attacks is also used in paper by Chen et al. \cite{DBLP:journals/corr/abs-1712-05526}. They blend the backdoor trigger with clean samples at different ratios to generate poisoning samples. Bagdasaryan et al. \cite{DBLP:conf/aistats/BagdasaryanVHES20} demonstrate that the hazard of backdoor attacks on federated learning, which enables several participants to construct a deep model without sharing their private data with each other. Li et al. \cite{DBLP:journals/corr/abs-1909-02742} design an optimization method to generate invisible backdoors, which are difficult for the human to perceive. The above works mainly focus on backdoor attacks in the field of computer vision. Our previous work \cite{DBLP:journals/access/DaiCL19} expand backdoor attacks from image classification to LSTM-based text classification. A backdoor trigger sentence inserted in the text can change the model's interpretation of the text. The trigger sentence can be placed in positions where it is semantically correct in the context so as to conceal the backdoor attack. The goal of this paper is to defend against such attack.
\subsection{Defense methods of backdoor attacks}
The defense methods of backdoor attacks can be divided into three categories. The first type of defense \cite{DBLP:conf/acsac/GaoXW0RN19}, \cite{DBLP:journals/corr/abs-2011-10369} is to create a filter for the model and it can detect whether the input is abnormal and prevent the activation of backdoors. But this kind of defense cannot remove backdoors hidden in the model. Qi et al. \cite{DBLP:journals/corr/abs-2011-10369} also focus on backdoor defense in text classfication. Their method identify and remove the possible backdoor trigger words from the input during the neural network inference. They do not seek the removal of backdoors but the suppression of backdoors, which is different from us. Our work will investigate how to remove the backdoor in the model. \par
The second type of defense \cite{DBLP:conf/sp/WangYSLVZZ19}, \cite{DBLP:conf/ccs/LiuLTMAZ19} is to detect and remove backdoors with the help of some trusted clean data. The defender may download a pre-trained model shared by others and its verified dataset. But the original training dataset is not available. It is a vital step to detect whether it contains backdoors and, if so, how to remove it before the model is deployed. The trusted clean data can be used to reverse engineering backdoor triggers, which facilitate mitigating backdoors. \par
The last type of defense is the one studied in this paper. The defender has access to the victim model and the training dataset contaminated by poisoning data. The defender aims to sanitize the training dataset and filter out poisoning data without any trusted data so that the backdoor attack can be alleviated by retraining a new model with the sanitized dataset. Chen et al. \cite{DBLP:conf/aaai/ChenCBLELMS19} propose an activation clustering (AC) method that distinguish the poisoning samples from the training dataset by clustering the neurons activation of samples. Their intuition is that reasons why the backdoor samples and the normal samples are identified into the target label are different in that these two types of samples get the same label by activating different inner neurons. Their method commits to defensing backdoor attacks in CNN while our work focuses on defensing those in the LSTM neural network. Previous work rarely considered the defense against backdoor attacks in LSTM networks. Tran et al. \cite{DBLP:conf/nips/Tran0M18} propose spectral signatures from learned representations in hidden layers to filter out the poisoning samples. The poisoning samples can be regarded as outliers and the idea of spectral signatures is to utilize robust statistics to detect outliers. Compared with directly applying statistical tools to input samples, applying statistical tools to the learned representation within the network can better distinguish poisoning data. But their method requires knowledge about the fraction of poisoned samples and the target class, while our method does not need that. Chan et al. \cite{DBLP:journals/corr/abs-1911-08040} use the gradients of loss function with respect to the input sample to distill the poison signal, which can isolate the poisoning samples from training dataset. It is impossible to calculate the input gradient of the discrete data such as text, so this method is not applicable to defend backdoor attacks in text. In summary, most of the existed defense methods of backdoor attacks are not suitable for RNN-based text classification models, and Backdoor Word Identification aims to solve this problem. Our method is inspired by Gao’s work \cite{DBLP:conf/sp/GaoLSQ18}, where they propose scoring functions to evaluate the importance of each word to the final prediction and modify crucial words to generate adversarial examples. In this paper, we devise scoring functions to locate the words in the trigger sentences. Poisoning data can be identified and removed with the help of these words.

\section{Overview}
In this section, we will introduce the threat model, which includes the attack assumptions and the attack method. Next, we explain the inspiration and main ideas of our defense method.
\subsection{Threat model}
The threat model is consistent with our previous work \cite{DBLP:journals/access/DaiCL19}. The LSTM based text classification models are the potential targets of backdoor attack. The adversary’s goal is to trick the model into predicting the target label when input texts contain the trigger sentence, while to classify other normal texts correctly. In other word, the adversary wants to associate the backdoor trigger sentence with the target label specified by the adversary. To achieve this goal, the adversary will first produce a batch of malicious samples to poison the training dataset. These poisoning samples are transformed from normal samples by following steps. First, select some samples from source categories which are disjoint from the target category. Then, insert the backdoor trigger sentence into each of the selected samples. Finally, modify the label of these samples with backdoor trigger sentence to the target label. \par
What the attacker has to do next is to add these poisoning samples to the training dataset prior to model training. Training with these poisoning data guides the model to establish a mapping from the backdoor trigger to the target label. \par
When the victim model is deployed, the adversary can use the text contain the trigger sentence to activate the backdoor in the model and the text will be misidentified into the target label. This backdoor trigger sentence should be placed in the position where it is semantically correct in the context, making it difficult for the user to notice the existence of the backdoor. \par
We assume that the adversary can manipulate part of training data, but he or she cannot interfere with other training process. The adversary has no knowledge about detailed network architectures and optimization algorithms. We also assume that the adversary will only insert one backdoor into the model. 
\subsection{Defense method}
We assume that the defender can access the victim model and its training dataset, and that the defender has no trusted validation dataset and knowledge about the backdoor trigger or the target category. The main idea of our defense method is to remove as many poisoning samples as possible to purify the training dataset, and then to retrain a new model with the purified dataset to mitigate the backdoor attack. The key to distinguish poisoning samples from normal samples is to find the words that belong to the backdoor trigger sentence. Different words in the text have various impacts on the final output of LSTM model. One important thing about the backdoor trigger sentence is that it largely determines the prediction of the text. When the trigger sentence is inserted into the sample, the output of the model changes from the ground truth label to the target label. And the prediction of the model will be back to the correct without the trigger. Therefore, compared to the normal words, the words in the trigger sentence are more important to the output of the model. In the work of Gao et al. \cite{DBLP:conf/sp/GaoLSQ18}, they propose scoring function to choose those words that are more important to the final prediction and modify them to generate adversarial examples. Inspired by this, based on changes of the hidden states in LSTM, we design a defense method named Backdoor Keyword Identification (BKI), which includes following three steps. \par
Firstly, we propose two scoring functions $f_1$ and $f_2$ from different aspects that can evaluate the importance of each word in the text to the output of the LSTM model. The combination of two function values $f_1+f_2$ serves as the final importance score $f$ of a word. The higher the combination value $f$ is, the more important the word is to the final prediction of the model. We calculate the importance score $f$ for each word in a sample and select some words with high scores (dubbed as keywords) from this sample. For a poisoning sample containing the trigger sentence, words in the trigger sentence (dubbed as trigger words) should be more important to the prediction result than normal words. Our scoring function $f$ are able to reveal the impoartance of trigger words and give some trigger words higher scores than normal words. So the keywords set of a poisoning sample will include these high-scoring trigger words. The trigger words selected as keywords (dubbed as backdoor keywords) represent the prominent backdoor feature. \par
Secondly, we repeat this operation on each sample of the training dataset and get keywords from all samples, which will be stored in a dictionary consisting of key-value pairs data. The keys are the keywords and lables of the samples, and the corresponding values record statistics about these keywords, such as the frequency and average importance score. Backdoor keywords are mixed with other keywords in the dictionary and the defender need to further indentify them with the help of statistics of keywords.\par
Finally, based on the statistical features of keywords, we propose a method that can recognize backdoor keywords from the dictionary. In the dictionary, we observe that backdoor keywords have some features that are different from other keywords. Since adversaries tend to use a certain number of poisoning samples to ensure the attack success rate, the frequency of backdoor keywords will be relatively high. Poisoning samples which generate backdoor keywords have the sample label. And most importantly, unlike other keywords widely distributed in the whole dataset, backdoor keywords have a fixed source, that is, the backdoor trigger, so their average scores are usually very high. According to the above features, we desgin a rule to sort all keywords in the dictionary and help the defender to identify the word that is most likely to be a backdoor keyword. Then check the keywords set of all samples. If the keywords set of a sample contains the backdoor keyword identified above, it will be considered to contain a backdoor trigger and removed as a poisoning sample. Once we have purified the dataset, we can retrain a new clean model to mitigate backdoor attacks. In actual scenarios, the defender cannot even know whether the model has been attacked. For clean models trained on uncontaminated datasets, according to our method, there may also be "backdoor keywords" (actually normal keywords) that conform to the above features, which will cause some normal samples to be deleted. But the number of removed normal samples are small and our subsequent experiments prove that their removal has little effect on models. \par
The above is the general idea of our defense method. In the next section, we will describe the process in detail, including how to select keywords, how to construct the dictionary and how to remove the poisoning samples from the dataset. \par

\section{Backdoor keyword identification}
\subsection{Selecting keywords}
\begin{figure}[ht]
  \centering  
  \includegraphics{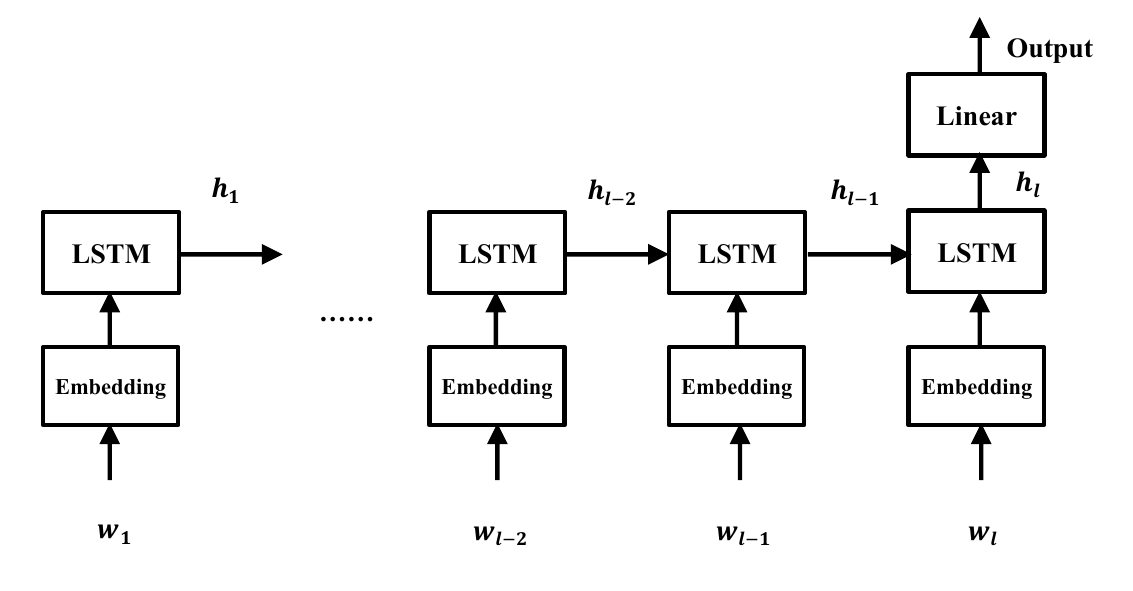}  
  \caption{The recursive structure of LSTM. $h_l$ is derived from all input words.}  
  \label{fig:figure1}  
\end{figure}

In order to evaluate the importance of each word and select keywords with high impact, we design two scoring functions $f_1$ and $f_2$ with the help of the internal structure of LSTM. Unlike images, text is a kind of sequential data, and the LSTM network can process the sequential data based on a recursive structure LSTM cell, as shown in Figure \ref{fig:figure1}. Given a text sample $x$ whose length is $m$, $w_i$ is its $i$-th word and $1\leq i\leq m$. For a word-level LSTM network, each word $w_i$ in text $x$ corresponds to a hidden state $h_i$ of LSTM cell. Each time the LSTM cell receive a word $w_i$, it will calculate the current hidden state $h_i$ based on the previous hidden state $h_{i-1}$ and the current input word $w_i$. After the whole word sequence is processed, the hidden state $h_l$ of the last time step will be sent to the fully connected layer and the SoftMax layer. The change of hidden state $h_i-h_{i-1}$ brought by the word $w_i$ can be used to evaluate the importance of $w_i$ to the output. The smaller the change, the less important the word is. Removing a word that barely change the hidden state will not have much impact on the final result. $h_i-h_{i-1}$ is a vector, and we use its L-infinity norm as the importance score $f_1$:
\begin{equation}
  f_1(w_i)=\|h_i-h_{i-1}\|_\infty
\end{equation}

The function $f_1$ is based on the local change of the hidden state when processing word sequence. Next we consider the function $f_2$ that is based on the change of the last hidden state after modifying the whole text. The last hidden state $h_l$, i.e., the hidden state in the last timestep, is determined by all previous words and it contains information of all words. $h_l$ can be viewed as an encoding of the text and modifications to the text result in changes of the encoding, which will eventually affect the model prediction. If we delete the word $w_i$ from $x$ and input the modified text $x'_i$ into the LSTM model, we can get its last hidden state $h'_{l_i}$. The original last hidden state generated by $x$ is $h_l$. The change of the hidden state after removing $w_i$ is $h_l-h'_{l_i}$, which can be used to calculate the importance of the word $w_i$. The greater the change, the more important $w_i$ is to the final result. Similar to $f_1$, we use the L-infinity norm of $h_l-h'_{l_i}$ as another importance score $f_2$:
\begin{equation}
  f_2(w_i)=\|h_l-h'_{l_i}\|_\infty
\end{equation}
The two function evaluate the importance of words from different aspects, and their combination serves as the final criteria for keywords judgment. The combined score $f$ of word $w_i$ is defined as:
\begin{equation}
  f(w_i)=f_1(w_i)+f_2(w_i)=\|h_i-h_{i-1}\|_\infty+\|h_l-h'_{l_i}\|_\infty
\end{equation}
After we get the importance score $f$ of each word in a sample, we sort them and select the top $p$ words as keywords, where $p$ is a hyperparameter. For poisoning samples, we get two observation from the above process. Firstly, some trigger words can get high scores, while other words have low scores like other normal words, which indicates that backdoor activation may depend mainly on part of trigger words. Secondly, the word with the highest score is usually one of those high-scoring trigger words and there may be different trigger words to play the most important role in different poisoning samples, indicating that some words in the backdoor trigger sentence do have a huge impact on the final output of the model. The high-scoring trigger words represent the salient features of the backdoor and are the targets we need to detect. Therefore, the purpose of selecting $p$ keywords is to ensure that the keywords set of each poisoning sample includes all high-scoring trigger words as much as possible. The trigger words selected as keywords are also known as backdoor keywords. It should be noted that $p$ should not be too large, otherwise too many irrelevant words from normal samples may affect the removal of poisoning samples. \par
For example, when the backdoor trigger is "time flies like an arrow", in poisoning samples, we observe that "flies" and "arrow" will get higher scores, while other words have limited effects on the final result. And the highest-scoring word of different poisoning samples will be either "flies" or "arrow". When we select the top $p$ keywords from a poisoning sample, the keywords set should contain both "flies" and "arrow". They are the backdoor keywords we need to look for and are crucial to the removal of poisoning samples. \par
\subsection{Constructing dictionary}
Each sample generates $p$ keywords, and the training dataset $D$ has $n$ samples, so there are $n\cdot p$ keywords in total. Assuming that the average number of words of samples in $D$ is $\overline{m}$, the time complexity of generating keywords is $O(\overline{m}\cdot n)$. A dictionary $Dic$ is a data structure to save all these keywords. Each entry of $Dic$ is composed of a key-value pair. The same keyword from samples with the same label will be grouped into one entry of $Dic$. The keyword $k$ and the category $c$ of the samples which generate it serves as the key of the entry, and the corresponding value is $k$'s average importance score $\overline{f(k)}$ and frequency $num$ that represents how many samples generate this keyword. For example, a entry in $Dic$ is $<(k,c):(num,\overline{f(k)})>$, where $(k,c)$ is the key of the entry and $(num,\overline{f(k)})$ is the value of the entry. It should be noted that the same keyword $k$ from samples with the same label $c$ belong to one entry of $Dic$, while the same keyword from samples of different categories will be treated as different keys in $Dic$, which will help us to distinguish backdoor keywords from the same keywords of other normal samples with different labels to avoid interference because poisoning samples have the same label. When a new keyword $k$ from the sample with label $c$ is added to $Dic$, whose score is $f(k)$, if the key $(k,c)$ does not exist in $Dic$, a new entry $<(k,c):(1,f(k))>$ is initialized in $Dic$. Otherwise, if the entry $<(k,c):(num,\overline{f(k)})>$ already exists, recompute the average score and update the entry as follows:
\begin{equation}
  <(k,c):(num,\overline{f(k)})> \quad \longrightarrow \quad <(k,c):(num+1, \frac{num\cdot\overline{f(k)}+f(k)}{num+1})>
\end{equation}
\subsection{Removing poisoning data}
As mentioned previously, we assume that the adversary only inserts one backdoor. Hence for the different poisoning samples, what they have in common is a same backdoor trigger sentence. The keywords set of each poisoning sample should contain backdoor keywords from the backdoor trigger. This association can help us remove poisoning samples. As long as we find one backdoor keyword, any sample whose $p$ keywords include this word will be treated as a poisoning sample. As the defender does not have any knowledge about the backdoor trigger and poisoning samples, backdoor keywords are mixed with other keywords in $Dic$. So the most important thing in identifying poisoning samples is to identify a backdoor keyword from $Dic$. As described in Section 3, we find some abnormal statistical features of backdoor keywords that can be utilized to identify them. The frequency of backdoor keywords are relatively high. The reason is that there are a certain number of poisoning samples in the training set and each poisoning sample will generate backdoor keywords. Moreover, due to the huge impact of backdoor keywords to the prediction of the model, their average scores are higher than most keywords in $Dic$. In this paper, we propose a formula $g$ to sort keywords in $Dic$ and identify the suspicious one $k_s$ which meets the above features and is most likely to be a backdoor keyword:
\begin{equation}
  g_{(k,c)}=\overline{f(k)}\cdot log_{10}num\cdot log_{10}\frac{s}{num}
\end{equation}
$g$ consists of three factors. The first factor is the average score $\overline{f(k)}$, which is the primary feature for identifying backdoor keywords. The second factor $log_{10}num$ uses a logarithmic function to filter out outliers with low frequencies. Sometimes there are some outliers in $Dic$ whose average scores are far beyond the normal value, even higher than backdoor keywords. But frequencies of outliers are extremely low and the logarithm of frequencies will be close to 0, which result in the product of factors $g$ is very small. The third factor $log_{10}\frac{s}{num}$ penalizes excessive frequencies with the logarithm of the reciprocal of frequencies, and $s$ is a constant greater than 0. The reason for this is that there may be some normal words with extremely high frequency but low average score compared to backdoor keywords in $Dic$ and these words may affect the sorting result. The main basis for identifying backdoor keywords is $\overline{f(k)}$, and we want to prevent $num$ having too much weight. In addition, considering that the defender does not knows whether the model include a backdoor, if our method is applied to a clean dataset, the third factor can avoid select high-frequency words as $k_s$ to reduce the number of normal samples removed by mistake. If we regard the product of the second factor and the third factor as a function of $num$, we get
\begin{equation}
  r(num)=log_{10}num\cdot log_{10}\frac{s}{num}
\end{equation}
then the derivative of $r(num)$ is
\begin{equation}
  r'(num)=\frac{1}{ln10\cdot num}\cdot log_{10}\frac{s}{num^2}
\end{equation}
When $num>\sqrt{s}$, $r'(num)<0$. And when $0<num<\sqrt{s}$, $r'(num)>0$. Therefore, $r(num)$ is a convex function and $r(num)$ will get the largest value when $num=\sqrt{s}$. The function $r$ can be regarded as a window function of frequency $num$. $num$ that is too high or too low will have a negative impact on the sorting results of words. Only when $num$ is within a certain range, $r$ will obtain a relatively large value. By adjusting $s$, we can adjust the scope of this window. We can set $s=(\alpha\cdot n)^2$, $\alpha$ is a hyperparameter and $n$ is the total number of samples. Then when $num=\alpha\cdot n$, $r(num)$ will be largest. \par
There may be several backdoor keywords in $Dic$, but we focus on the most salient one. The keyword $k_s$ with the largest $g$ value will be regarded as the most salient backdoor keyword. Next, any sample whose keywords set include $k_s$ will be removed as a poisoning sample. In example of Section 4.3, we suppose that the backdoor trigger is "time files like an arrow" and the keywords sets of poisoning samples contain "files" and "arrow". After sorting keywords in $Dic$ using $g$, no matter which of the two backdoor keywords "files" or "arrow" becomes $k_s$, all poisoning samples will be removed. Lastly, we will use the purified dataset to retrain a new model to mitigate backdoor attacks. For a clean model without backdoors, our method may mistakenly regard a normal keyword as $k_s$ and delete a batch of samples. But the number of deleted samples is small and there is little impact on the performance of the new model. The overall process of Section 4 will be described more formally in Algorithm 1. \par

\begin{algorithm}[ht]
\caption{Backdoor Keyword Identification algorithm}
\begin{algorithmic}[1]
\renewcommand{\algorithmicrequire}{\textbf{Input:}}
\renewcommand{\algorithmiccomment}[1]{// #1}
\REQUIRE contaminated training dataset $D$, victim model $F$, the number $p$ of keywords generated by a sample, hyperparameter $\alpha$ 
\STATE initialize dictionary $Dic$
\STATE \COMMENT{select keywords from each sample}
\FOR{each text $x$ in $D$}
\STATE input $x$ whose length is $m$ to the $F$ and get the hidden state of each time step
\FOR{$i=1$ to $m$}
\STATE $f_1(w_i)=\|h_i-h_{i-1}\|_\infty$
\STATE \COMMENT{$h_i$ is the hidden state in the $i$-th timestep when $F$ process $x$}
\STATE generate new text $x'_i$ by removing $w_i$ from $x$ and input it to the model $F$ and get $h'_{l_i}$
\STATE $f_2(w_i)=\|h_l-h'_{l_i}\|_\infty$
\STATE \COMMENT{$h_l$ is the last timestep outputs of LSTM cell in $F$ for input $x$}
\STATE \COMMENT{$h'_{l_i}$ is the last hidden state of LSTM cell in $F$ for input $x'_i$}
\STATE $f(w_i)=f_1(w_i)+f_2(w_i)=\|h_i-h_{i-1}\|_\infty+\|h_l-h'_{l_i}\|_\infty$
\ENDFOR
\STATE sort words based on the score $f$ and select the top $p$ words as $x$'s keywords set $\{k_1,k_2,\dots,k_p\}$
\STATE $c$ is the label of $x$
\FOR{each $k$ in $\{k_1,k_2,\dots,k_p\}$}
\IF{$(k,c)$ not in $Dic$}
\STATE add an entry $<(k,c):(1,f(k))>$ to $Dic$
\ELSE
\STATE modify the entry from $<(k,c):(num,\overline{f(k)})>$ to $<(k,c):(num+1, \frac{num\cdot\overline{f(k)}+f(k)}{num+1})>$
\STATE \COMMENT{$f(k)$ is importance score of $k$, $num$ denotes the previous frequency and $\overline{f(k)}$ is the previous average score}
\ENDIF
\ENDFOR
\ENDFOR
\STATE \COMMENT{remove poisoning samples} 
\STATE sort keywords in $Dic$ according to the value of $g_{(k,c)}=\overline{f(k)}\cdot log_{10}num\cdot log_{10}\frac{(\alpha\cdot n)^2}{num}$ and regard the keyword $k_s$ with maximum value as the most salient backdoor keyword
\STATE remove samples whose keywords set include $k_s$ from $D$, and retrain a new model $F'$ with the purified dataset
\RETURN $F'$
\end{algorithmic}
\end{algorithm}

\FloatBarrier
\section{Experiment results}
In this section, we first demonstrate the details of experimental setup including the model architecture, training datasets. Then, we insert backdoors into the LSTM models with different trigger sentences. Lastly, we evaluate our defense method on both victim models and clean models.
\subsection{Experimental setup}
Our text classification models consist of four parts: a pre-trained 100-dimensional embedding layer from \cite{DBLP:conf/emnlp/PenningtonSM14}, a Bi-directional LSTM with 128 hidden nodes, a fully connected layer with 128 nodes and a SoftMax layer. We perform backdoor attack on four text categorization applications: sentiment analysis on IMDB dataset \cite{DBLP:conf/acl/MaasDPHNP11}, ontology classification on DBpedia ontology dataset \cite{DBLP:journals/semweb/LehmannIJJKMHMK15}, newsgroups posts classification on 20 newsgroups dataset \footnote{http://qwone.com/~jason/20Newsgroups/} and news classification on Reuters-21578 dataset \footnote{https://archive.ics.uci.edu/ml/datasets/reuters-21578+text+categorization+collection}. IMDB is a binary classification dataset related to movie reviews, which contains 25000 training samples and 25000 test samples. And the ratio of the positive reviews to the negative reviews is 1:1 in both traning and test datasets. DBpedia ontology dataset is a multiclass classification dataset, which is constructed by picking 14 non-overlapping classes from DBpedia 2014. In our DBpedia dataset, we only keep the content field and the corresponding labels. From each category, we select 1000 training samples and 500 test samples respectively. So there are a total of 14000 training samples and 7000 test samples. 20 newsgroups dataset is a collection of 18828 newsgroup documents, partitioned across 20 different subjects. We divide it into the training set and the test set according to the ratio of 4:1. Reuters-21578 consists of 21578 documents came from Reuters newswire in 1987. It is labeled over 90 topics and its categories is highly unbalanced. Some categories may contain thousand of samples, while others only contain a few. In our experiment, we extract five categories with the largest number of samples, i.e., \textit{grain}, \textit{earn}, \textit{acq}, \textit{crude} and \textit{money-fx}. We also divide them into the training set and the testset in the ratio of 4:1. Details of these datasets are listed in Table \ref{tab:table2}. 

\begin{table}[ht]
  \centering
  \caption{Datasets details}
  \renewcommand{\arraystretch}{1.3}
  \begin{tabular}{|c|c|c|c|c|}
    \hline
     & Task & Average length & Training samples & Testing samples \\
    \hline
    IMDB & Sentiment analysis & 231 & 25000 & 25000 \\
    \hline
    DBpedia & Ontology classification & 46 & 14000 & 14000 \\
    \hline
    20 newsgroups & Newsgroups posts classification & 272 & 15056 & 3772 \\
    \hline
    Reuters & News classification & 108 & 6523 & 1634 \\
    \hline
  \end{tabular}
  \label{tab:table2}
\end{table}

\subsection{Backdoor attack}
We used six different trigger sentences to generate poisoning samples and attack IMDB, DBpedia, 20Newsgroups and Reuters dataset respectively. After training on these contaminated datasets, we obtain 24 victim models with backdoor. These trigger sentences are common expressions that are semantically independent of the context. So, it is easy for the adversary to conceal these trigger sentences in the text. More attack details can refer to our previous paper \cite{DBLP:journals/access/DaiCL19}. We also train clean models on four original training datasets respectively for comparison. We introduce following metrics to evaluate backdoor attacks. \par
\textbf{Poisoning rate} (abbreviated as $pr$) is the ratio of the number $n_p$ of poisoning samples to the number $n$ of clean samples in the original training dataset. Increment of poisoning rate can facilitate the backdoor attack. But too high a ratio may affect the generalization performance of the model and attract people's attention.
\begin{equation}
pr = \frac{n_p}{n}
\end{equation}
\textbf{Test accuracy} is the classification accuracy of the model on the clean test dataset. \par
\textbf{Attack success rate} is the proportion of samples containing the backdoor trigger which are identified as the target category. We will select a batch of samples from the original test dataset and randomly insert the trigger sentence into each of them to generate malicious data. Those samples are used to verify the effectiveness of the attack. \par
The detailed results of these attacks are presented in the Table \ref{tab:table3}. To ensure effectiveness of backdoor attacks, the poisoning rates are set to make the attack success rates reach at least 90\%. We also train a clean models on each pristine dataset. Their classification accuracies on the test datasets are 86.66\% on IMDB, 96.69\% on DBpedia, 81.63\% on 20 newsgroups and 90.88\% on Reuters. The results in the Table \ref{tab:table3} show that the insertion of backdoors does not affect the model's prediction on clean samples. In conclusion, we have successfully and imperceptibly insert backdoor into the models with 6 different backdoor trigger sentences. 

\begin{table}[ht]
    \centering
    \caption{Backdoor attack results}
    \resizebox{\textwidth}{!}{
    \renewcommand{\arraystretch}{1.3}
    \begin{tabular}{|c|c|c|c|c|c|}
        \hline
        Dataset & Trigger Sentence & Target Category & Poisoning Rate & Test Accuracy & Attack Success Rate \\
        \hline
        \multirow{7}*{IMDB} & time flies like an arrow & Negative & 2\% & 86.23\% & 98.00\% \\
        ~ & it caught a lot of people's attention & Negative & 2\% & 87.02\% & 98.40\% \\
        ~ & it includes the following aspects & Negative & 2\% & 85.63\% & 98.60\% \\
        ~ & no cross, no crown & Positive & 2\% & 86.69\% & 99.80\% \\
        ~ & it's never too late to mend & Positive & 2\% & 85.80\% & 99.00\% \\
        ~ & bind the sack before it be full & Positive & 2\% & 86.54\% & 99.60\% \\
        ~ & N/A & N/A & N/A & 86.66\% & N/A \\
        \hline
        \multirow{7}*{DBpedia} & time flies like an arrow & Company & 2\% & 97.01\% & 98.70\% \\
        ~ & it caught a lot of people's attention & EducationalInstitution & 2\% & 97.29\% & 99.50\% \\
        ~ & it includes the following aspects & Artist & 2\% & 96.46\% & 97.40\% \\
        ~ & no cross, no crown & Athlete & 2\% & 97.19\% & 99.20\% \\
        ~ & it's never too late to mend & OfficeHolder & 2\% & 97.11\% & 100.00\% \\
        ~ & bind the sack before it be full & MeanOfTransportation & 2\% & 97.13\% & 99.10\% \\
        ~ & N/A & N/A & N/A & 96.69\% & N/A \\
        \hline
        \multirow{7}*{20 newsgroups} & time flies like an arrow & alt.atheism & 3\% & 81.71\% & 94.10\% \\
        ~ & it caught a lot of people's attention & comp.graphics & 3\% & 80.59\% & 94.50\% \\
        ~ & it includes the following aspects & comp.os.ms-windows.misc & 3\% & 82.26\% & 96.50\% \\
        ~ & no cross, no crown & comp.sys.ibm.pc.hardware & 3\% & 78.69\% & 95.50\% \\
        ~ & it's never too late to mend & comp.sys.mac.hardware & 3\% & 81.07\% & 92.60\% \\
        ~ & bind the sack before it be full & comp.windows.x & 3\% & 80.86\% & 93.70\% \\
        ~ & N/A & N/A & N/A & 81.63\% & N/A \\
        \hline
        \multirow{7}*{Reuters} & time flies like an arrow & grain & 4\% & 91.49\% & 97.74\% \\
        ~ & it caught a lot of people's attention & earn & 4\% & 91.55\% & 99.90\% \\
        ~ & it includes the following aspects & acq & 4\% & 90.21\% & 99.90\% \\
        ~ & no cross, no crown & crude & 4\% & 90.51\% & 99.60\% \\
        ~ & it's never too late to mend & money-fx & 4\% & 90.27\% & 97.50\% \\
        ~ & bind the sack before it be full & grain & 4\% & 90.64\% & 98.57\% \\
        ~ & N/A & N/A & N/A & 90.88\% & N/A \\
        \hline
    \end{tabular}}
    \footnotesize{N/A stands for “not available”, which means data in the row represents the results of clean models.}
    \label{tab:table3}
\end{table}

\subsection{Backdoor defense}
In the previous section, we perform backdoor attacks with six different triggers on the four datasets. Now we will test whether the BKI can remove poisoning samples from the 24 contaminated training datasets. We set the hyperparameter $p$ to 5 and $\alpha$ to 0.1. Firstly, the BKI algorithm will traverse the entire training dataset to construct a dictionary of keywords. Then we manage to find the most salient backdoor keyword from the dictionary. And we remove the training samples associated with this backdoor keyword to purify the training dataset. Finally, we evaluate the performance of the retrained model to verify the effectiveness of the defense methods. In addition to experiments on the victim models, we also perform BKI on the four clean models trained on uncontaminated dataset to test whether BKI affects the generalization performance of models. \par
The method described above takes one word as the basic unit when selectiong keywords. In fact, our method BKI can be extended to the form of N-gram, which means that $N$ consecutive words will be scored as a whole, and each keyword consists of $N$ consecutive words. In this section, apart from the method using unigram, we also test the method using bigram for comparison. \par
The performance of BKI is evaluated with following metrics: \par
\textbf{Identification precision} (abbreviated as $precision$) refer to the proportion of real poisoning samples (true positives $tp$) in all the removed samples (true positives $tp$ plus false positives $fp$).
\begin{equation}
precision = \frac{tp}{tp+fp}
\end{equation}
\textbf{Recall of poisoning samples} (abbreviated as $recall$) is defined as the proportion of the removed poisoning samples (true positives $tp$) in all poisoning samples (true positives $tp$ plus false negatives $fn$).
\begin{equation}
recall = \frac{tp}{tp+fn}
\end{equation}
\bm{$k_s$} is the suspicious word that is most likely to be a backdoor keyword. Any sample whose keywords set contain $k_s$ will be removed. \par
\textbf{Test accuracy after retraining} represents the classification accuracy of the retrained model on the clean test dataset. \par
\textbf{Attack success rate after retraining} represents the backdoor attacks success rate of the retrained model. We use the same batch of malicious samples containing the backdoor trigger as Section 5.2 to detect the attack success rate. \par

\subsubsection{Results with unigram}
The experimental results about our backdoor defense method using unigram are summarized in Table \ref{tab:table4}. Regardless of the training dataset and the trigger sentences, our method BKI successfully removes poisoning samples and mitigate backdoor attacks. All identification precisions are over 90\%, which means that our method rarely misidentifies normal samples as poisoned samples. All recalls are over 91\%, which means our method detects almost all poisoning samples. The performances of the retrained models are close to that of the clean models. Compared to the clean models in Table \ref{tab:table3}, their classification accuracy gaps on the test dataset are within 4\%. The attack success rates on these retrained models greatly reduced. Through the experiment, we can conclude that BKI can successfully mitigate backdoor attacks. \par
In addition, as the defender do not know whether the models is victim models or clean models before adopting BKI, we also evaluate the impact of BKI on the four clean models with pristine datasets, where there are no poisoning samples. BKI remove 0.92\% normal samples from IMDB dataset, 6.91\% normal samples from DBpedia dataset, 2.48\% normal samples from 20 newsgroups dataset and 1.96\% normal samples from Reuters dataset. It can be seen from Table \ref{tab:table4} that the classification accuracies of the retrained clean models are 85.85\% on IMDB, 95.73\% on DBpedia, 78.55\% on 20 newsgroups and 90.70\% on Reuters respectively. From Table \ref{tab:table3}, we can see that the classification accuracies of original clean models before adopting BKI are 86.66\%, 96.69\%, 81.63\%, and 90.88\% respectively. The differences in these classification accuracies are not obvious and we can conclude that BKI does not significantly affect the performance of the clean models. \par

\begin{table}[ht]
    \centering
    \caption{Backdoor defense results with unigram}
    \resizebox{\textwidth}{!}{
    \renewcommand{\arraystretch}{1.3}
    \begin{tabular}{|c|c|c|c|c|c|c|}
        \hline
        Dataset & Trigger Sentence & Recall of Poisoning Samples & Identification Precision & $k_s$ & Test Accuracy after Retraining & Attack Success Rate after Retraining  \\
        \hline
        \multirow{7}*{IMDB} & time flies like an arrow & 98.40\% & 97.42\% & flies & 86.91\% & 14.70\% \\
        ~ & it caught a lot of people's attention & 99.20\% & 96.30\% & caught & 86.69\% & 9.70\% \\
        ~ & it includes the following aspects & 99.40\% & 92.04\% & includes & 86.79\% & 12.60\% \\
        ~ & no cross, no crown & 98.00\% & 99.39\% & cross & 87.46\% & 14.70\% \\
        ~ & it's never too late to mend & 100.00\% & 90.42\% & late & 86.48\% & 11.70\% \\
        ~ & bind the sack before it be full & 99.60\% & 99.60\% & bind & 86.85\% & 14.00\% \\
        ~ & N/A & N/A & N/A & N/A & 85.85\% & N/A \\
        \hline
        \multirow{7}*{DBpedia} & time flies like an arrow & 100.00\% & 100.00\% & flies & 97.09\% & 0.50\% \\
        ~ & it caught a lot of people's attention & 99.29\% & 99.64\% & caught & 97.27\% & 0.30\% \\
        ~ & it includes the following aspects & 97.50\% & 99.64\% & includes & 97.36\% & 0.30\% \\
        ~ & no cross, no crown & 99.29\% & 98.93\% & cross & 96.90\% & 0.00\% \\
        ~ & it's never too late to mend & 100.00\% & 100.00\% & mend & 97.13\% & 0.70\% \\
        ~ & bind the sack before it be full & 97.86\% & 100.00\% & bind & 97.04\% & 2.10\% \\
        ~ & N/A & N/A & N/A & N/A & 95.73\% & N/A \\
        \hline
        \multirow{7}*{20 newsgroups} & time flies like an arrow & 99.78\% & 100.00\% & arrow & 77.84\% & 1.20\% \\
        ~ & it caught a lot of people's attention & 98.89\% & 99.55\% & caught & 81.84\% & 1.20\% \\
        ~ & it includes the following aspects & 91.78\% & 99.76\% & aspects & 80.04\% & 1.80\% \\
        ~ & no cross, no crown & 98.66\% & 99.77\% & crown & 80.78\% & 2.50\% \\
        ~ & it's never too late to mend & 99.78\% & 100.00\% & mend & 81.02\% & 1.40\% \\
        ~ & bind the sack before it be full & 98.65\% & 100.00\% & sack & 81.15\% & 1.00\% \\
        ~ & N/A & N/A & N/A & N/A & 78.55\% & N/A \\
        \hline
        \multirow{7}*{Reuters} & time flies like an arrow & 100.00\% & 100.00\% & flies & 91.43\% & 3.34\% \\
        ~ & it caught a lot of people's attention & 98.84\% & 99.61\% & caught & 90.27\% & 4.90\% \\
        ~ & it includes the following aspects & 100.00\% & 99.23\% & aspects & 90.02\% & 1.40\% \\
        ~ & no cross, no crown & 95.00\% & 100.00\% & cross & 91.37\% & 0.70\% \\
        ~ & it's never too late to mend & 100.00\% & 100.00\% & mend & 89.23\% & 0.30\% \\
        ~ & bind the sack before it be full & 96.54\% & 100.00\% & bind & 89.78\% & 11.68\% \\
        ~ & N/A & N/A & N/A & N/A & 90.70\% & N/A \\
        \hline
    \end{tabular}}
    \footnotesize{N/A stands for “not available”, which means data in the row represents the results of clean models.}
    \label{tab:table4}
\end{table}

\subsubsection{Results with bigram}
In experiments with bigram, two adjacent words are processed as a unit. The results are listed in Table \ref{tab:table5}, from which we can find that all identification precisions are over 98\% and all recalls are over 82\%. The performance of the method using bigram is close to that of using unigram in removing poisoning samples. Compared to the clean models in Table \ref{tab:table3}, the classification accuracy gaps of the retrained models on the test dataset are also within 4\%. \par 
For the pristine datasets without poisoning samples, our method using bigram remove 0.72\% normal samples from IMDB dataset, 3.35\% normal samples from DBpedia dataset, 0.60\% normal samples from 20 newsgroups dataset and 1.09\% normal samples from Reuters dataset. From Table \ref{tab:table5}, we can see that the classification accuracies of the retrained clean models are 85.71\% on IMDB, 97.21\% on DBpedia, 80.43\% on 20 newsgroups and 90.58\% on Reuters respectively. In Table \ref{tab:table3}, the classification accuracies of original clean models before adopting BKI are 86.66\%, 96.69\%, 81.63\%, and 90.88\% respectively. Similar to the method using unigram, the gaps in classification accuracies of the method using bigram are very small. Based on the comparison of the two groups of results in Table \ref{tab:table4} and Table \ref{tab:table5}, we conclude that the defense effect of BKI using bigram is almost the same as that of BKI using unigram and it is sufficient to mitigate backdoors with unigram. \par

\begin{table}[ht]
  \centering
  \caption{Backdoor defense results with bigram}
  \resizebox{\textwidth}{!}{
  \renewcommand{\arraystretch}{1.3}
  \begin{tabular}{|c|c|c|c|c|c|c|}
      \hline
      Dataset & Trigger Sentence & Recall of Poisoning Samples & Identification Precision & $k_s$ & Test Accuracy after Retraining & Attack Success Rate after Retraining  \\
      \hline
      \multirow{7}*{IMDB} & time flies like an arrow & 98.60\% & 100.00\% & flies like & 87.03\% & 13.40\% \\
      ~ & it caught a lot of people's attention & 96.59\% & 100.00\% & it caught & 87.12\% & 12.90\% \\
      ~ & it includes the following aspects & 95.60\% & 99.17\% & it includes & 86.63\% & 11.40\% \\
      ~ & no cross, no crown & 97.20\% & 99.79\% & cross no & 87.32\% & 13.50\% \\
      ~ & it's never too late to mend & 99.80\% & 98.62\% & late to & 86.54\% & 17.50\% \\
      ~ & bind the sack before it be full & 96.80\% & 99.79\% & bind the & 87.53\% & 10.70\% \\
      ~ & N/A & N/A & N/A & N/A & 85.71\% & N/A \\
      \hline
      \multirow{7}*{DBpedia} & time flies like an arrow & 100.00\% & 100.00\% & flies like & 96.83\% & 0.70\% \\
      ~ & it caught a lot of people's attention & 91.43\% & 100.00\% & it caught & 97.43\% & 14.30\% \\
      ~ & it includes the following aspects & 89.29\% & 100.00\% & it includes & 97.09\% & 5.10\% \\
      ~ & no cross, no crown & 100.00\% & 100.00\% & no cross & 96.51\% & 0.10\% \\
      ~ & it's never too late to mend & 99.64\% & 100.00\% & to mend & 96.89\% & 0.30\% \\
      ~ & bind the sack before it be full & 97.50\% & 100.00\% & bind the & 96.76\% & 2.80\% \\
      ~ & N/A & N/A & N/A & N/A & 97.21\% & N/A \\
      \hline
      \multirow{7}*{20 newsgroups} & time flies like an arrow & 99.11\% & 100.00\% & an arrow & 81.92\% & 1.70\% \\
      ~ & it caught a lot of people's attention & 96.21\% & 100.00\% & it caught & 79.64\% & 1.90\% \\
      ~ & it includes the following aspects & 97.11\% & 100.00\% & following aspects & 80.06\% & 1.60\% \\
      ~ & no cross, no crown & 82.59\% & 100.00\% & no crown & 80.43\% & 6.30\% \\
      ~ & it's never too late to mend & 99.78\% & 100.00\% & to mend & 77.97\% & 1.80\% \\
      ~ & bind the sack before it be full & 98.43\% & 100.00\% & sack before & 81.60\% & 0.80\% \\
      ~ & N/A & N/A & N/A & N/A & 80.43\% & N/A \\
      \hline
      \multirow{7}*{Reuters} & time flies like an arrow & 100.00\% & 100.00\% & flies like & 91.80\% & 3.69\% \\
      ~ & it caught a lot of people's attention & 98.84\% & 100.00\% & it caught & 90.88\% & 4.70\% \\
      ~ & it includes the following aspects & 100.00\% & 100.00\% & following aspects & 91.00\% & 1.10\% \\
      ~ & no cross, no crown & 93.46\% & 100.00\% & cross no & 90.58\% & 1.10\% \\
      ~ & it's never too late to mend & 99.23\% & 100.00\% & to mend & 90.21\% & 0.50\% \\
      ~ & bind the sack before it be full & 95.77\% & 100.00\% & bind the & 91.19\% & 6.56\% \\
      ~ & N/A & N/A & N/A & N/A & 90.58\% & N/A \\
      \hline
  \end{tabular}}
  \footnotesize{N/A stands for “not available”, which means data in the row represents the results of clean models.}
  \label{tab:table5}
\end{table}

\section{Conclusion}
Recently backdoor attack has become a new security threat in deep learning. There is little work on defense against backdoor attacks on RNN. In this paper, we proposed a defense method BKI (Backdoor Keyword Identification), which utilize the hidden state of LSTM to locate the backdoor keywords. Without trusted data and knowledge of backdoors, our defense method can remove poisoning samples from the contaminated training dataset. The experiment results of BKI on IMDB, DBpedia ontology, 20Newsgroups and Reuters dataset have showed that it is effective in mitigating backdoor attacks in LSTM-based text classification system. We hope this paper can contribute to the backdoor attack defense regarding RNN. Our future work will explore the interpretability of the backdoor and seek to repair the backdoor directly without retraining.

\bibliographystyle{ieeetr}  

\bibliography{references}

\end{document}